# Comprehensive TCAD Simulation Study of High Voltage (>650V) Common Drain Bidirectional AlGaN/GaN HEMTs

Md Tahmidul Alam and Chirag Gupta

*Abstract*— A broad TCAD simulation analysis of a monolithic common drain bidirectional GaN HEMT was performed. We used gate-to-gate distances of 4μm and 6μm for the devices optimized with two field plates. The breakdown voltages were 675V and 915V respectively. Inclusion of field plates near both the gates produced electric field peaks at the opposite ends of the transistor simultaneously. This resulted in better electric field management or higher blocking voltage per unit length. Consequently, the 675V monolithic bidirectional HEMT had an impressive 40% improvement in on-resistance than its 650V typical series/parallel counterpart.

## I. INTRODUCTION

In some novel power electronic topologies, such as AC-AC matrix converter [1], multilevel T-type inverter [2] and DC-DC dual active bridge (DAB) [3], it is required to allow and block current in both directions. To achieve bidirectional operation, traditionally two unidirectional transistors are arranged in anti-series [4] or anti-parallel [5] configuration. However, these designs are not ideal because they double the device count and have three contact electrodes. As a result, it causes ~2x on-resistance, increased volume, higher cost, higher conduction loss, higher complexity, increased form-factor and slower transient response. Monolithic bidirectional switch (MBDS) implemented by GaN HEMTs is highly promising to reduce these problems since it has the potential to lower the on-resistance with the same bidirectional voltage blocking capability [6].

The fundamental structure of a monolithic bidirectional HEMT is composed of the typical AlGaN/GaN layers of a conventional GaN HEMT (Fig. 1). It has two ohmic sources ($S_1$ and $S_2$) and two gates ($G_1$ and $G_2$) near each source. Each gate is biased with respect to its nearest source to make the operation symmetric and hence stable [7]. The substrate is kept floating to get a symmetric on-resistance [8]. There are several reports on the structure [9], [10] however detailed device structure description such as field plates is lacking. Additionally, extensive semiconductor device level operation and quantitative understanding of four quadrant switches (FQS) or MBDS is not readily available in the literature and only one or two studies have focused on this aspect [11]. In this paper, we have tried to address these issues by performing a comprehensive TCAD simulation study to understand the transistor's behavior and obtain device parameters. In our design, we have utilized two field plates close to each gate to manage the peak electric field and to maximize the breakdown voltage.

Our simulations indicate that the inclusion of field plates into this structure has a distinctive property of accumulating secondary electric field peaks at the field plates near the applied positive bias ($S_2$). This fact further increases voltage sharing along the channel. Consequently, our simulation results demonstrate an attractive ~40% improvement in on-resistance compared to typical series/parallel switches type implementation of bidirectional transistors.

## II. DEVICE STRUCTURE AND OPERATION

The device shown in Fig.1 was simulated for two gate-to-gate lengths, 4μm and 6μm in Silvaco TCAD. A 25nm epitaxial AlGaN layer with 30% Al composition was put on top of a 1.5μm UID GaN layer followed by a 3 μm semi-insulating UID GaN substrate. Two sources, two gates and four field plates were integrated into the structure. Each field plate was shorted to its nearest source. We used two field plates for both the structures to keep the fabrication and design complexities the same. The field plate positions were adjusted to have almost equal electric field peaks at a horizontal cutline through AlGaN (Fig. 2). The device was made symmetrical to get a symmetric electrical response. $Si_3N_4$ was used as the dielectric medium and passivation layer. Detailed structural description of the device is mentioned in Table I.

Since the closest field plates, $FP_3$ and $FP_4$ were connected to the opposite polarities of the applied voltage, a substantial gap was kept between them to prevent dielectric breakdown. Assuming the dielectric strength of $Si_3N_4$ to be 9MV/cm [12], $FP_3$ and $FP_4$ were kept 1.5μm apart so that the electric field never exceeded 4.5 MV/cm.

The threshold voltage of this structure is -6V. In order to release electron from $S_1$ to the 2DEG channel at the AlGaN/GaN interface, $V_{g1s1}$ should be ≥ -6V. Similarly, $V_{g2s2}$ should be ≥-6V to release electrons from $S_2$ to the channel. By modulating the gate voltages, it is possible to allow and block current in both directions. It is also possible to allow and block current in one direction that makes this device suitable for versatile applications.

Both spontaneous and piezoelectric sheet charges were added to our simulation models. The piezoelectric charge was calculated by the stress created by the lattice mismatch between AlGaN and GaN. Silvaco's standard models for low-field and high-field mobility of GaN were used for calculating mobility in varying electric fields. Schottky-Read-Hall recombination model and Selberherr's impact ionization model were used to mimic electron-hole recombination and avalanche effect at breakdown [13].

TABLE I
DEVICE STRUCTURE DESCRIPTION

| Parameter | Description | Value |
|---|---|---|
| $L_G$ | Gate length | 1μm |
| $L_{OHM}$ | Source/Drain Length | 0.5μm |
| $L_{FP}$ | Field Plate Length | 0.5μm |
| $L_{GS}$ | Gate to Source Distance | 250nm |
| $L_{GF}$ | Gate to Field Plate Distance | 100nm |
| $L_{FF}$ | Field Plate to Field Plate Distance | 100nm |
| $T_{FP1}$ | Dielectric thickness under first field plate | 50nm |
| $T_{FP2}$ | Dielectric thickness under second field plate | 200nm |

This work was supported by Materials Research Science and Engineering Center (MRSEC), University of Wisconsin-Madison, WI 53706, United States of America.

M.T. Alam and C. Gupta are with the Department of Electrical and Computer Engineering, University of Wisconsin-Madison, WI 53706, United States of America (email: malam9@wisc.edu).

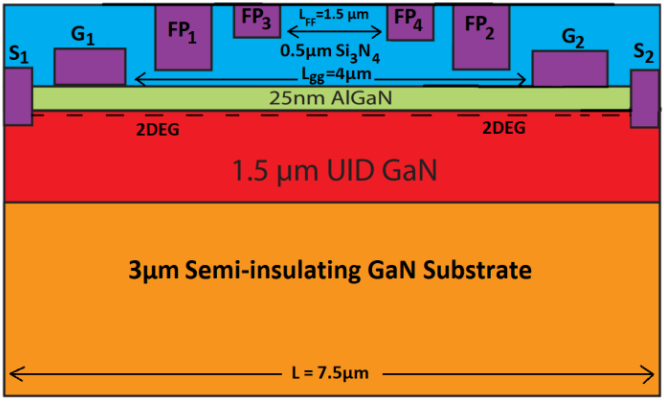

Fig.1. Structure of the proposed device with $L_{gg}$ = 4μm.

### III. RESULTS AND DISCUSSIONS

*A. DC Response*

For the gate-to-gate lengths of 4μm and 6μm the breakdown voltages of the MBDS were 675V and 915V respectively (Fig. 3). Due to the inclusion of field plates on both sides, three secondary field peaks accumulated at the gate and field plates near the applied positive bias ($S_2$) as shown in Fig. 2. To investigate whether this property results in higher blocking voltage sharing along the channel than unidirectional HEMT's and to quantitatively compare their performance, we implemented two unidirectional HEMTs of almost similar breakdown voltages (650V and 850V). Both were optimized with two field plates. The required channel length (4μm) of the 675V MBDS was 12.5% lower than the channel length (4.5μm) of the 650V unidirectional HEMT. It was challenging to obtain the breakdown voltage of the unidirectional HEMT above 800V with two field plates even with >9μm channel length. Adding a 3$^{rd}$ field plate might have been effective, however it would increase the design complexies. Therefore, an 825V unidirectional HEMT with channel length of 9.75um was taken as the representative of unidirectional HEMT for comparison with the 915V MBDS.

If both the gates of the MBDS are on or off simultaneously, the device operates in bidirectional mode (Fig. 4). Fig. 5 shows the IV curves of the device in unidirectional mode, where one gate is always on ($V_{g2s2} = 0$) and the other gate is off with different gate voltages ($V_{g1s1}$ = -8V, -10V, -12V and -14V). Therefore, the device allows current only from $S_1$ to $S_2$. One interesting property of the unidirectional IV curves is that current flows through the device only when $V_{g1} - V_{S2} \geq -6V$. It indicates that for unidirectional operation $G_1$ acts as the gate and $S_2$ acts as the source with the same threshold voltage as bidirectional mode. Hence the offset voltage for conduction is always equal to $V_{g1s1}$ plus 6V. The 675V MBDS had an on-resistance of 4.44 Ω.mm ($V_{g1s1} = V_{g2s2} = -3V$, $V_{DS} = 2V$) whereas the on-resistance of the 650V unidirectional device was 3.7 Ω.mm under similar test conditions. Since the on resistance of a typical series/parallel switch would be ~2x3.7 Ω.mm, the MBDS showed a tremendous 40% improvement.

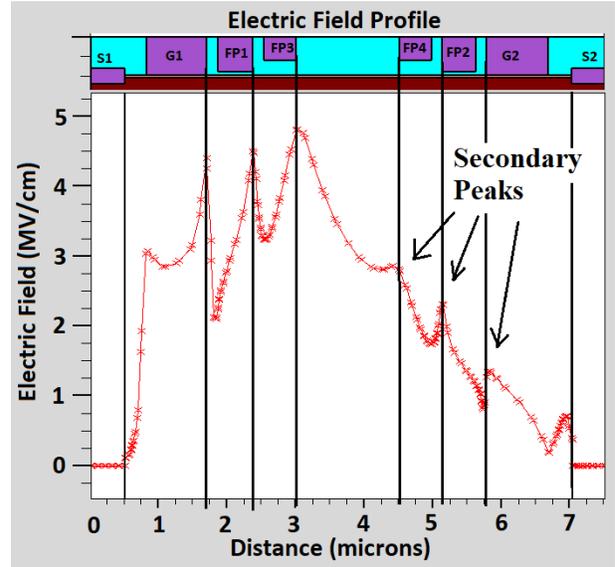

Fig.2. Electric field profile along AlGaN cutline at the onset of breakdown, for $V_{S2} > V_{S1}$ and $L_{gg}$ = 4μm. Indicating extra peaks near the positive bias.

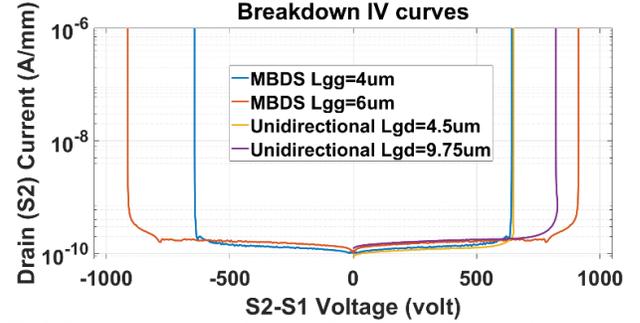

Fig.3. Breakdown voltages for both MBDS and unidirectional switches.

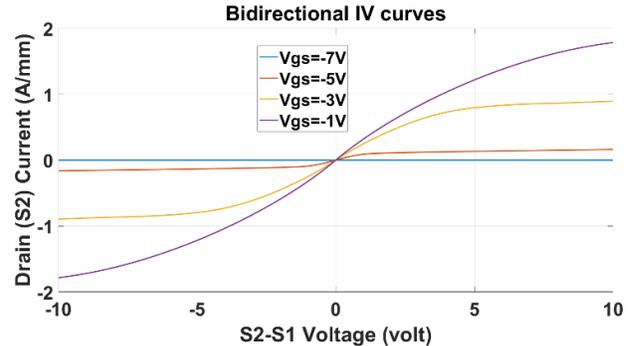

Fig.4. Bidirectional IV curves, for $L_{gg}$ = 4μm, $V_{g1s1} = V_{g2s2}$.

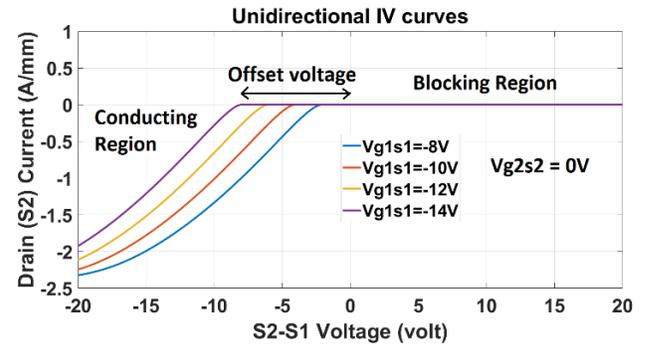

Fig.5. Unidirectional mode IV curves for $L_{gg}$ = 4μm, $V_{g2s2} = 0$ and $V_{g1s1} < -6V$.

*B. AC response*

The off-state capacitance of the output node (S2) is plotted against the applied terminal voltage in Fig. 6. COSS falls rapidly at the beginning then remains practically constant. The initial quick fall is due to the depletion of the 2DEG with applied bias. For drain voltage <20V, the unidirectional devices have slightly higher capacitance than their bidirectional counterparts. This may be due to the shielding of the output node (S2) of the MBDS by the gate and field plates near S2.

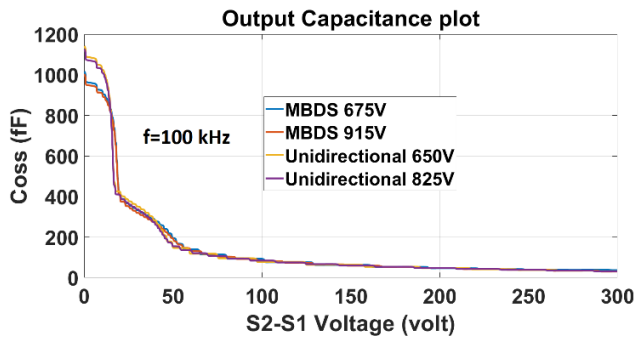

Fig.6. Output Capacitance for both MBDS and unidirectional switches.

*C. Transient Reponse*

The transient response of the MBDS was found by applying a step input into the gate from 0V to -15V while keeping the terminal voltages at a constant potential of 5V (Fig. 7). The 915V MBDS had slower response than the 675V MBDS because RON increased linearly with channel length, but COSS remained reasonably constant. Therefore, the time constant (product of RON and COSS) increased. This indicates that the transient performance of HEMT structure is dominated by the on-resistance hence the appreciable reduction of on-resistance in monolithic bidirectional HEMTs can lead to a significant improvement in the operational speed of the system.

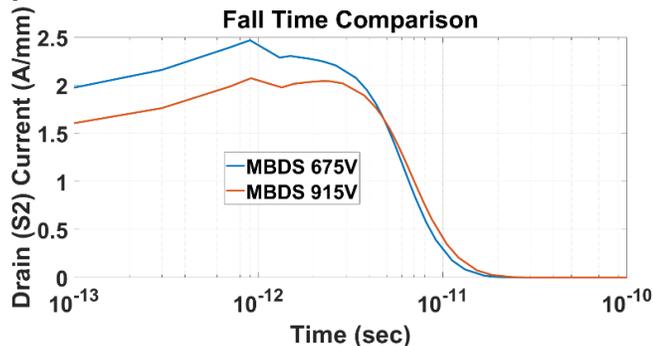

Fig.7. Fall time for a step input for the 675V and 915V MBDS.

A summary of the key properties of the 675V, 4μm MBDS is listed in Table II.

TABLE II

| Parameter | Value | Test Conditions |
|---|---|---|
| Breakdown voltage | 675V | $V_{g1s1} = V_{g2s2} = -15V$ |
| On-resistance | 4.44 Ω.mm | $V_{g1s1} = V_{g2s2} = -3V$, $V_{S1-S2} = 2V$ |
| C$_{OSS}$ | 278 fF | $V_{g1s1} = -15V$, $V_{S1-S2} = 40V$ |
| Fall Time | 10.9 ps | Time to reach 10% of initial value |

## IV. CONCLUSION

We presented a detailed TCAD simulation analysis of a monolithic bidirectional AlGaN/GaN HEMT with field plates to obtain deep insights of its electrical behavior and device parameters to fill up an existing knowledge gap. Its performance was quantitively compared with typical anti-series and/or anti-parallel bidirectional HEMT's. We observed a unique peak electric field presence at the opposite sides of the device simultaneously due to field plate integration in both sides. Consequently, this structure had higher blocking voltage per channel length compared to unidirectional HEMTs. As a result, there was a remarkable ~40% fall in on-resistance compared to traditional anti-series/anti-parallel bidirectional HEMTs for the same breakdown voltage.